\documentclass[twocolumn, floatfix, nofootinbib,superscriptaddress]{revtex4-1}

\usepackage{amsmath}
\usepackage{amsthm}
\usepackage{amsfonts}
\usepackage{graphicx}
\usepackage{tikz}
\usepackage{mathtools}
\usepackage{physics}
\usepackage{bm}
\allowdisplaybreaks
\usetikzlibrary{automata,positioning}
\usepackage{empheq}
\usepackage{dsfont}
\usepackage{tensor}
\usepackage{color}
\usepackage{xcolor}

\definecolor{niceBlue}{RGB}{20,10,237}
\usepackage[colorlinks,citecolor=niceBlue,linkcolor=niceBlue,urlcolor=niceBlue,hyperindex]{hyperref}

\newcommand{\aind}{i}
\newcommand{\bind}{j}
\newcommand{\cind}{k}

\begin{document}

\title{Dynamical Axions in \texorpdfstring{$\bm{U(1)}$}{U(1)} Quantum Spin Liquids}

\author{Salvatore D. Pace}
\affiliation{Department of Physics, Massachusetts Institute of Technology, Cambridge, Massachusetts 02139, USA}
\affiliation{TCM Group, Cavendish Laboratory, University of Cambridge, Cambridge CB3 0HE, United Kingdom}

\author{Claudio Castelnovo}
\affiliation{TCM Group, Cavendish Laboratory, University of Cambridge, Cambridge CB3 0HE, United Kingdom}

\author{Chris R. Laumann}
\affiliation{Department of Physics, Boston University, Boston, Massachusetts 02215, USA}

\date{\today}

\begin{abstract}
Since their proposal nearly half a century ago, physicists have sought axions in both high energy and condensed matter settings. Despite intense and growing efforts, to date experimental success has been limited, with the most prominent results arising in the context of topological insulators. 
Here we propose a novel mechanism whereby axions can be realized in quantum spin liquids. 
We discuss the necessary symmetry requirements and identify possible experimental realizations in candidate pyrochlore materials.
In this context, the axions couple \textit{both} to the external and to the emergent electromagnetic fields. We show that the interaction between the axion and the emergent photon leads to a characteristic dynamical response, which can be measured experimentally in inelastic neutron scattering. 
This work sets the stage for studying axion electrodynamics in the highly tunable setting of frustrated magnets. 
\end{abstract}

\maketitle

Quantum spin liquids (QSL) are long-range entangled phases of matter with fractionalized spins and emergent gauge fields~\cite{anderson1987resonating, savary2016quantum,zhou2017quantum}. 
One of the most fascinating examples is a $U(1)$ QSL, where the gauge structure resembles quantum electrodynamics~\cite{levin2005colloquium, henley2010coulomb}. 
Vastly different from our universe, this emergent quantum electrodynamics violates Lorentz symmetry, is strongly coupled, and contains magnetic monopoles. 
Such $U(1)$ QSLs have attracted substantial interest due to proposed experimental realizations~\cite{hermele2004pyrochlore,Shannon2012,gingras2014quantum} in a class of frustrated quantum rare-earth pyrochlore magnets called quantum spin ice (QSI)~\cite{PhysRevX.1.021002,PhysRevLett.115.097202, pan2016measure, PhysRevB.94.144415, sibille2018experimental, gaudet2019quantum, gao2019experimental}.

The long-wavelength effective description of a $U(1)$ QSL in 3+1D with only gapped matter is in terms of an emergent $U(1)$ gauge field whose quanta are photons, electric gauge charges corresponding to fractionalized spin excitations (spinons), and dual magnetic monopoles (visons)~\cite{ftTerm}.
At energies below the matter gap, the universal properties of $U(1)$ QSL phases stem from their emergent photons governed by the Maxwell Lagrangian ${\mathcal{L}_{\gamma} = \left(|\bm{e}|^{2} - |\bm{b}|^{2}\right)/2}$, where $\bm{e}$ and $\bm{b}$ are \textit{emergent} electric and magnetic fields (see SM for our choice of electromagnetic units~\cite{supRef}). 

A tantalizing modification to the Maxwell Lagrangian that has been entertained both in high energy and condensed matter physics is the axion term ${\mathcal{L}_{\varphi\gamma} = \alpha \varphi \left(\bm{e}\cdot\bm{b}\right)/\pi}$~\cite{wilczek1987two}, where ${\alpha = e^{2}/(4\pi c)}$ is the fine structure constant\footnote{Here $c$ is the `speed of light' and $e$ is the elementary `electric charge', both associated with the emergent gauge structure.}~\cite{pace2020emergent} and $\varphi$ is a real pseudoscalar called the axion field. 
In high-energy physics, axions are considered one of the best motivated particles beyond the standard model, acting as a remedy to the strong $CP$ problem~\cite{RevModPhys.82.557}, naturally arising in string theory~\cite{svrcek2006axions}, and playing a role as a possible dark-matter candidate~\cite{duffy2009axions}. 
Despite decades of intensive experimental efforts, such axions have not been observed~\cite{graham2015experimental}.

Axions can also emerge as collective excitations in condensed matter systems~\cite{anderson1972more, wilczek1987two}. 
Indeed, the fluctuations of any (anti)ferromagnetic ordering with a pseudoscalar order parameter couple as a dynamical axion field at long wavelengths. 
In the context of topological insulators, a spacetime constant $\varphi$ (in which case, the axion term is called a $\theta$-term) plays an important role in the electromagnetic response~\cite{PhysRevB.78.195424, cho2011topological, chan2013effective}, and the influence of dynamical axionic fluctuations on the external electromagnetic fields has been studied in considerable detail~\cite{li2010dynamical, wu2016quantized, nenno2020axion, sekine2021axion}.
Additionally, they have been discussed in the context of topological superconductors~\cite{PhysRevB.87.134519} as well as Weyl semimetals~\cite{PhysRevB.87.161107,gooth2019axionic}.

\begin{figure}[t!]
\centering
    \includegraphics[width=.48\textwidth]{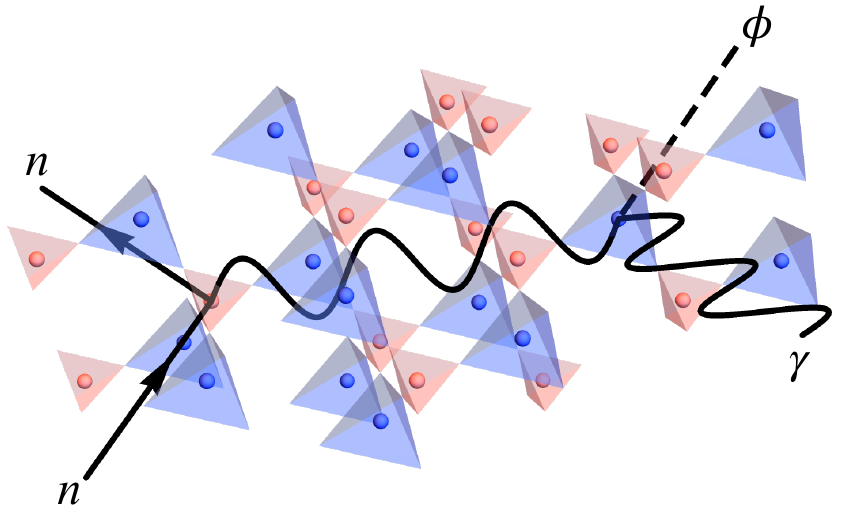}
    \caption{
    In the breathing pyrochlore lattice, the A (red) and B (blue) tetrahedra have different volume, breaking inversion symmetry. 
    A staggered spinon charge density, with positive (red sphere) and negative (blue sphere) spinons on A and B tetrahedra, respectively, breaks time-reversal~\cite{footnoteTRSbreaking}. 
    The combined order couples as an axion to the internal electrodynamics. 
    The embedded Feynman diagram illustrates one of the neutron scattering processes brought about by photon-axion pair production.
}
    \label{fig:pyrocartoon}
\end{figure}

Here we investigate the effects of an emergent dynamical axion in $U(1)$ QSLs.
This scenario has received limited attention in the literature and remains poorly understood beyond some work on the effect of $\theta$-terms~\cite{pesin2010mott,cho2012dyon,wang2016time,pretko2017higher}.
Like in topological insulators, the dynamical axion we study couples to external electromagnetic fields.
However, it also couples to the emergent electrodynamics, giving rise to a vacuum with elementary photons and axions that has been hitherto inaccessible in other contexts. 
We study this internal axion electrodynamics by investigating the dynamical structure factor, and show how it leads to prominent signatures accessible through inelastic neutron scattering. 
Namely, we find a characteristic two-particle continuum associated with the threshold production of axion-photon pairs. 
We further discuss possible experimental realizations in QSI on the breathing pyrochlore lattice, where an appropriate magnetic order interacts with the emergent gauge field as an axion field, see Fig.~\ref{fig:pyrocartoon}. 
This identifies breathing pyrochlore QSL candidates, like ${\text{Ba}_{3}\text{Yb}_{2}\text{Zn}_{5}\text{O}_{11}}$~\cite{kimura2014experimental}, as a possible condensed-matter realization of emergent axion electrodynamics.


\textit{Low-energy effective action} ---
$U(1)$ QSLs are described by the deconfined phase of $U(1)$ lattice gauge theory~\cite{chandrasekharan1997quantum}. 
This is well established in microscopic spin models~\cite{PhysRevLett.88.011602, PhysRevLett.89.277004, PhysRevB.68.184512,hermele2004pyrochlore}. 
Regularizing the axion term on a lattice is a long-standing open problem in lattice gauge theory, and current proposals involve complicated, beyond nearest-neighbor interactions~\cite{cardy1982phase,figueroa2018lattice, sulejmanpasic2019abelian, PW220703544}.

We hereby leave these technicalities, which do not affect the universal physics, for future work and we focus instead on coarse-graining the microscopic degrees of freedom and writing down the leading symmetry-allowed terms in a long-wavelength effective action~\cite{supRef}. 
We separate the axion field into a vacuum expectation value and fluctuations: ${\varphi(\bm{x},t) \equiv \theta + \phi(\bm{x},t)}$.
This leads to the low-energy effective action ${S_{\mathrm{eff}} = \int \mathrm{d}t~\mathrm{d}^{3}\bm{x}~\left(\mathcal{L}_{\gamma} + \mathcal{L}_{\phi} + \mathcal{L}_{\varphi\gamma}\right)}$, where 
\begin{subequations}\label{eqn:action}
\begin{align}
\mathcal{L}_{\gamma} &= \frac{1}{2}\left(|\bm{e}|^{2} - |\bm{b}|^{2}\right),\\
\mathcal{L}_{\phi} &=\frac{J}{2}\left(\left( \partial_{t}\phi\right)^{2} - v^{2}\left|\grad\phi\right|^{2} - \Delta^{2}\phi^{2} \right),\\
\label{eqn:Laxion}\mathcal{L}_{\varphi\gamma} &= \frac{\alpha}{\pi} \left(\theta + \phi\right) \bm{e}\cdot\bm{b}
\, 
\end{align}
\end{subequations}
(See SM~\cite{supRef} for additional discussion of ${S_{\mathrm{eff}}}$'s construction).
$\mathcal{L}_{\gamma}$ and $\mathcal{L}_{\varphi\gamma}$ are the Maxwell and axion terms, governing the emergent electric and magnetic fields $\bm{e}(\bm{x},t)$ and $\bm{b}(\bm{x},t)$ with ${\alpha\sim 1/10}$ the emergent fine structure constant~\cite{pace2020emergent}. 
$\mathcal{L}_{\phi}$ is the free theory of $\varphi$, where the phenomenological parameters $J$, $v$, and $\Delta$ correspond to the axion's stiffness, asymptotic speed, and gap, respectively. 
Except when $v = c$, the emergent speed of light, the axion sector of the theory violates the emergent Lorentz invariance.

In this work, we do not consider effects from the dynamical gapped electric gauge charges (spinons) and magnetic monopoles (visons) in $S_{\text{eff}}$.
Futhermore, $S_{\text{eff}}$ could include terms with external electric and magnetic fields $\bm{E}(\bm{x},t)$ and $\bm{B}(\bm{x},t)$ interacting with $\varphi$ (i.e., ${\mathcal{L}_{\text{ext}}\sim\varphi\bm{E}\cdot\bm{B}}$). 
Here, we focus on how the axion interacts with the emergent electromagnetic fields, Eq.~\eqref{eqn:Laxion}, as this physics in new to the QSL context.
We return briefly to the coupling to external electromagnetic fields at the end of this letter.

It is well known that the $\theta$-term is a total derivative and does not affect the dynamical equations of motion in the bulk. 
However, the presence of the axion term modifies the emergent Gauss's law as a Gauss-Witten law~\cite{FISCHLER1983165}
\begin{equation}\label{eqn:axionGaussLaw}
    \grad\cdot\bm{e} = - \frac{\alpha}{\pi}\grad\cdot(\varphi\bm{b})
    \, .
\end{equation}
This reveals the Witten effect, where the axion field causes magnetic monopoles to accumulate electric charge~\cite{witten1979dyons}.
For our purposes, it will also be important in identifying the degrees of freedom to which external probe-fields couple.


%
%
\begin{figure*}[t!]
\centering
    \includegraphics[width=\textwidth]{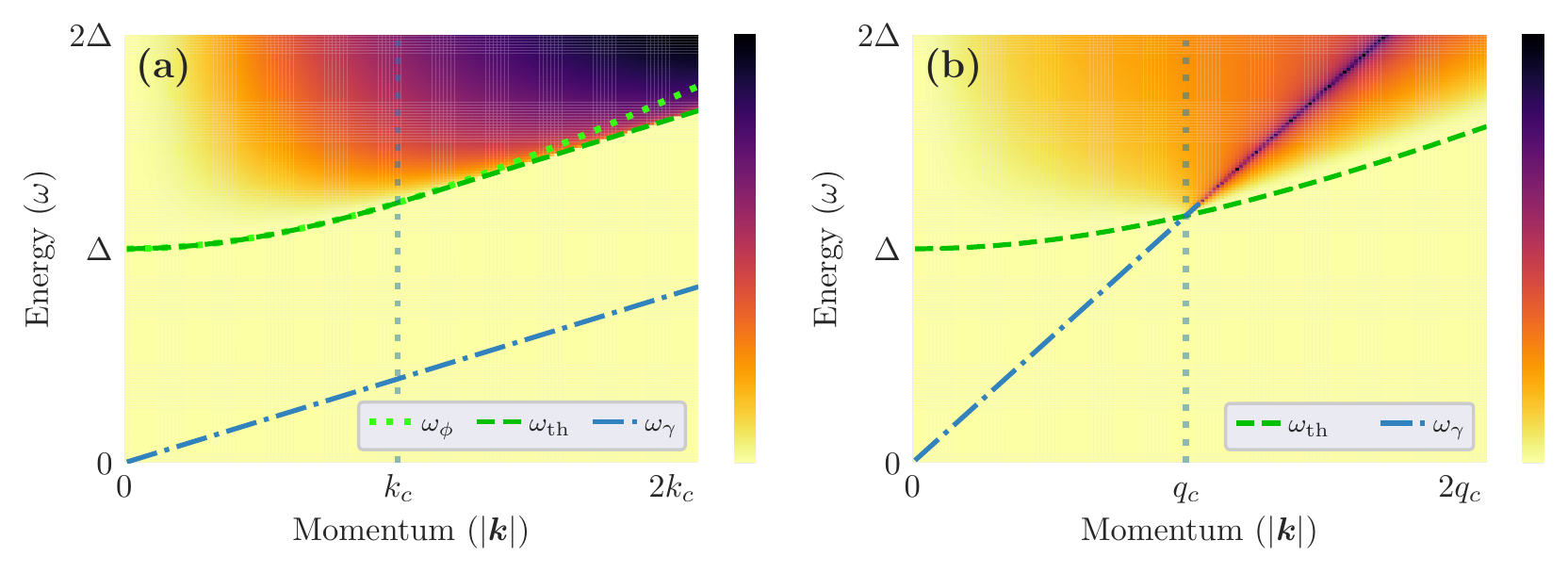}
    \caption{
    The dynamical structure factor $\mathcal{F}^{aa}(\bm{k},\omega)$ exhibits a two-particle continuum from axion-photon pair production above a threshold $\omega_{\mathrm{th}}$ (green dashed line, see Eq.~\eqref{eqn:NSFdymSF}). 
    Below threshold, the delta function response due to single photon states is indicated by $\omega_{\gamma}$ (blue dash-dotted line).
    Depending on the kinematics of the axion dispersion, either the axion can travel faster than the speed of light (a) or not (b).
    In the former case, the critical momentum beyond which the axion would travel faster than the speed of light is $k_c$. In the latter case, the single-photon dispersion enters the photon-axion continuum above momentum $q_c$,
    leading to a resonance centered at $\omega_{\gamma}$ (vertical dotted lines in the respective panels). 
    }
    \label{fig:heatmap}
\end{figure*}

\textit{Axions in quantum spin ice} --- 
Before turning to our results on the dynamics of axion-photon production, we contextualize them to the case of QSI. 
This is both because QSI has become a familiar model system for $U(1)$ QSLs~\cite{hermele2004pyrochlore,Shannon2012,gingras2014quantum}, and also because of its direct experimental relevance~\cite{PhysRevX.1.021002,PhysRevLett.115.097202, pan2016measure, PhysRevB.94.144415, sibille2018experimental, gaudet2019quantum, gao2019experimental}. 

We need to identify an experimentally motivated order parameter, $\varphi$, that interacts with the emergent photon as an axion.
To be consistent with the field-theoretic modeling above, upon coarse-graining, $\varphi$ must be a local time-reversal odd pseudoscalar to linearly couple with $\bm{e}\cdot\bm{b}$.

It is tempting to consider a common magnetic order parameter in pyrochlore systems~\cite{castelnovo2012spin}, namely
\begin{equation}\label{eqn:Qop}
Q_{t} = (-1)^{t} \operatorname{div}_{t}S^{(z)}
\, .
\end{equation}
Here, $t$ labels tetrahedra on the pyrochlore lattice and ${(-1)^{t} = \pm 1}$ depending on the sublattice ($A$ or $B$).
$\operatorname{div}_{t}S^{(z)}$ is the lattice divergence of $S^{(z)}$ --- the component of the spin along the local easy axis --- which yields the spinon charge at $t$.
Evidently, $Q_{t}$ is an order parameter for a staggered background of spinons, which corresponds to the spins developing all-in-all-out order and spontaneously breaking time-reversal symmetry~\cite{footnoteTRSbreaking}.
However, $Q_t$ is even under inversion and does not have the correct symmetry properties.

We are therefore brought to consider the breathing pyrochlore lattice, where $A$ and $B$ tetrahedra are of unequal size and thus no longer relate under inversion (see Fig.~\ref{fig:pyrocartoon}).
For small enough breathing anisotropy, the $U(1)$ QSL phase of QSI remains stable~\cite{savary2016quantumBreath}. 
The breathing anisotropy only modifies the microscopic ring-exchange scale, changing the value of the internal speed of light but not the stability of the effective field theory.

Breathing anisotropy is characterized by the lattice order parameter
\begin{equation}\label{eqn:deltaVop}
\Delta V_{l} = a^{-3}\left(V_{l}^{(A)} - V_{l}^{(B)}\right)
\, ,
\end{equation}
where, $V_{l}^{(A)}$ $(V_{l}^{(B)})$ is the volume of the $A$ ($B$) tetrahedron with corner at site $l$, and $a$ is the lattice constant. Evidently, $\Delta V$ is a time-reversal even real pseudoscalar, and the product $Q \Delta V$ finally has the correct symmetry. 
Coarse-graining and decomposing the order parameters into their vacuum expectation values and amplitude fluctuations, we obtain an axion field
\begin{equation}\label{eqn:axionQSI}
\varphi(\bm{x},t) \equiv \theta + \phi(\bm{x},t) \sim \langle\Delta V\rangle\langle Q\rangle + \langle\Delta V\rangle\delta Q(\bm{x},t)
\, .
\end{equation}
Here we have neglected fluctuations of $\Delta V$ since lattice distortions generally are higher energy than the magnetic fluctuations.

The dynamics of QSI at energies below the spinon gap is governed by a six-spin ring-exchange term~\cite{hermele2004pyrochlore}. Since $\phi$ and $e$ arise from single-body operators while $b$ arises from a six-body operator, the lattice axion term is an eight-body sub-leading ring-exchange term. 
Given that a six-body term is important to the dynamics of the QSL phase, an eight-body term may be sub-leading, but not negligible.


\textit{Axion-photon production} ---
To probe the coupled dynamics of the axion-photon system, we study the dynamical structure factor (DSF) of the coarse-grained spin magnetic moment $\bm{S}(\bm{x},t)$. 
This can be measured by neutron scattering and is often used to probe candidate spin liquids~\cite{knolle2019field, gao2019experimental, PhysRevX.1.021002, PhysRevB.94.144415, gaudet2019quantum}.
The DSF $\mathcal{F}^{\aind\bind}(\bm{k},\omega)$ is given by the imaginary part of the spin susceptibility associated with the correlation function ${\langle S^{\aind}(\bm{k},\omega) S^{\bind}(-\bm{k},-\omega)\rangle}$, where $\aind,\bind=x,y,z$.

The most interesting contribution to the DSF within the low-energy effective theory, Eq.~\eqref{eqn:action}, arises from the threshold production of axion-photon pairs.
While we do not model spinon-axion interactions, we note that they could modify the DSF in the spin-flip channel and cause an additional multi-particle continuum~\cite{huang2018dynamics, morampudi2020spectroscopy}.

To compute the DSF, we must connect $\bm{S}(\bm{x},t)$ to the degrees of freedom present in the effective field theory.
Below the spinon gap, physical states satisfy the divergence-free condition $\grad\cdot\bm{S} = 0$.
Comparing this to the Gauss-Witten law, Eq.~\eqref{eqn:axionGaussLaw}, up to a non-universal constant, $\bm{S}(\bm{x},t)$ is therefore given by
\begin{equation}\label{eqn:SzAxion}
\bm{S}(\bm{x},t)  = \bm{e}(\bm{x},t) + \frac{\alpha}{\pi}\varphi(\bm{x},t) \bm{b}(\bm{x},t)
\, .
\end{equation}
Eq.~\eqref{eqn:SzAxion} can be further confirmed by noting that the emergent vector potential and spin magnetic moment are canonically conjugate~\cite{hermele2004pyrochlore, savary2012coulombic, PhysRevB.95.134439}.
Identifying the coarse-grained $\bm{S}(\bm{x},t)$ as the canonical momentum of the effective field theory, the variation $\delta S_{\mathrm{eff}}/\delta\bm{e}$ again yields Eq.~\eqref{eqn:SzAxion}.

With this identification, the DSF $\mathcal{F}^{\aind\bind}(\bm{k},\omega)$ must be transverse.  
Solving the Schwinger-Dyson equation to leading loop order, we find (see SM~\cite{supRef}),
\begin{equation}\label{eqn:NSFdymSF}
\mathcal{F}^{\aind\bind} \hspace{-1pt}=\hspace{-1pt} \left(\delta^{\aind\bind} - \frac{ k^{\aind} k^{\bind}}{k^2} \right)\hspace{-1pt}
\begin{cases}
   \mathcal{F}_{\gamma}(\bm{k},\omega)\delta\left(\omega - c|\bm{k}|\right) & \omega < \omega_{\mathrm{th}}(\bm{k}) \\ 
   \mathcal{F}_{\mathrm{\phi\gamma}}(\bm{k},\omega) & \omega > \omega_{\mathrm{th}}(\bm{k}).
\end{cases}
\end{equation}
Here, $\omega_{\mathrm{th}}$ indicates the threshold energy above which axion-photon pairs can be produced at total momentum $\bm{k}$. 
The associated two-particle continuum is represented by $\mathcal{F}_{\mathrm{\phi\gamma}}$, which in the absence of the emergent axion vanishes.
Below the threshold, only single photon states are available and the DSF contains a delta function reflecting the photon dispersion, weighted by the function ${\mathcal{F}_{\gamma} = \left(\pi  + \alpha^{2}\theta^{2}/\pi \right) \omega}$. 
This reduces to the well known result for standard $U(1)$ QSLs when $\theta=0$~\cite{savary2012coulombic, benton2012seeing}.

\begin{figure}[t]
\centering
    \includegraphics[width=.48\textwidth]{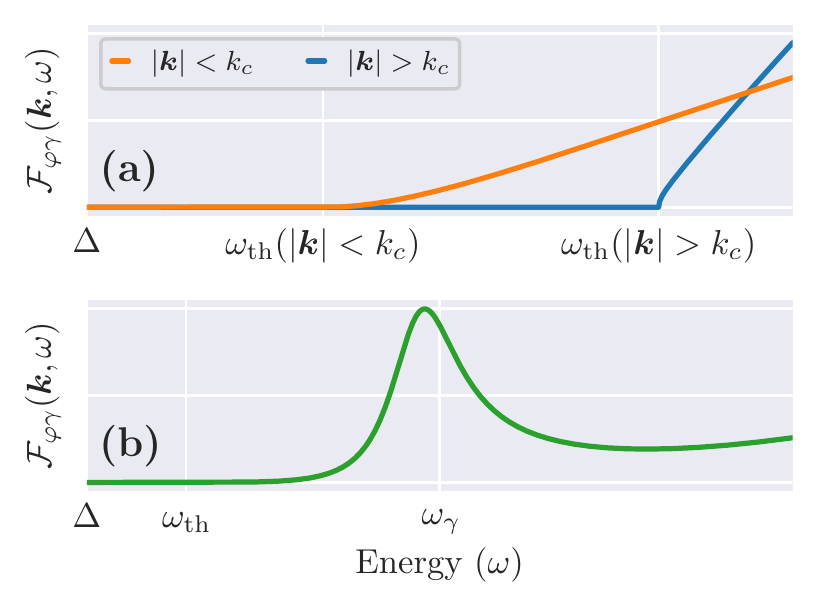}
    \caption{
    Line cuts of the dynamical structure factor $\mathcal{F}_{\mathrm{\phi\gamma}}(\bm{k},\omega)$ at fixed momentum $|\bm{k}|$ illustrating, for different kinematic regimes, (a) the turn on of the axion-photon continuum above threshold and (b) the single photon resonance on top of the continuum.
    For convenience, we have chosen parameters that produce a broad resonance with visible Fano asymmetry on the scale of the background.
    }
    \label{fig:kSlice}
\end{figure}

In general, the effective theory is not Lorentz invariant unless the asymptotic velocity of the axion field, $v$, equals the speed of light, $c$.
Thus, there are two distinct kinematic regimes for axion-photon production corresponding to $v\geq c$ and $v<c$, as illustrated in Fig.~\ref{fig:heatmap}(a) and (b) respectively.
In the case $v\geq c$, there is a critical momentum ${k_{c} = \Delta c/(v\sqrt{v^{2}-c^{2}})}$ beyond which the axion travels faster than $c$.
This leads to the threshold
\begin{equation}\label{eqn:NSFomegaTh}
\omega_{\mathrm{th}} = 
\begin{cases}
\omega_{\phi}\left(\bm{k}\right) \quad &|\bm{k}| < k_{c}\\
\omega_{\gamma}\left(|\bm{k}|-k_{c}\right) + \omega_{\phi}\left(k_{c}\right) \quad &|\bm{k}| \geq k_{c}
\end{cases} 
\, ,
\end{equation}
where ${\omega_{\phi}\left(|\bm{k}|\right) = \sqrt{\Delta^{2} + v^{2}|\bm{k}|^{2}}}$ and ${\omega_{\gamma}\left(|\bm{k}|\right) = c|\bm{k}|}$ are the axion and photon dispersions, respectively.
Physically, for external momenta less than $k_{c}$, all of the momentum is carried by the axion at threshold. 
For greater momenta, the axion travels at $c$ and the rest of the momentum is carried by the photon. 
While the full functional form of $\mathcal{F}_{\mathrm{\phi\gamma}}$ is cumbersome (see SM~\cite{supRef}), the behavior just above threshold is governed by characteristic exponents,
\begin{equation}\label{eqn:fThBehavoir}
\mathcal{F}_{\phi\gamma}(\bm{k},\omega) \sim
\begin{cases}
\left(\omega-\omega_{\mathrm{th}}\right)^{3} \quad &|\bm{k}| \leq k_{c}\\
\sqrt{\omega-\omega_{\mathrm{th}}}\quad  \quad &|\bm{k}| > k_{c}\\
\end{cases}
\, .
\end{equation}
See Fig.~\ref{fig:kSlice}(a) for frequency dependent line cuts of the full DSF illustrating these two behaviors. 
Notice that the naive expectation, based on the density of axion-photon states, goes as ${(\omega-\omega_{\mathrm{th}})^{2}}$ for $|\bm{k}|<k_{c}$ and ${\sqrt{\omega-\omega_{\mathrm{th}}}}$ for $|\bm{k}|>k_{c}$. 
The difference can be traced to the momentum dependence of the interaction.

In the case that $v<c$, the axion never reaches the speed of light so $\omega_{\mathrm{th}} = \omega_{\phi}$.
However, the single-photon dispersion crosses above $\omega_{\mathrm{th}}$ for momenta greater than $q_{c} = \Delta/\sqrt{c^{2}-v^{2}}$. 
In this regime, a single photon can decay into an axion-photon pair and the DSF develops a resonant peak above threshold, see Fig.~\ref{fig:kSlice}(b).
The Fano-like asymmetry~\cite{PhysRev.124.1866} of the resonant line-shape can be traced to the interference between the $\bm{e}$ operator in Eq.~\eqref{eqn:SzAxion} creating a single-photon state which decays into an axion-photon pair and the $\varphi \bm{b}$ operator directly producing such a pair. 

The experimental signatures of the axion discussed thus far are from interactions with photons.
It is interesting to consider whether a dynamical axion field can be detected through inelastic neutron scattering in the absence of an emergent electrodynamics.
In this case, there are no long-wavelength pseudovectors at the $\Gamma$ point to couple to the neutron spin.
Returning to the example of breathing pyrochlore, this indicates that fluctuations of $Q$ remain hidden to neutrons unless there is coexisting $U(1)$ fractionalization.


\textit{Coupling to external electromagnetic fields} ---
On symmetry grounds, the order parameter $\phi \sim Q \Delta V$ couples to the external electromagnetic field, leading to a magneto-electric polarizability in linear response.
Since the charge gap is large in these materials, the response of the conduction and valence bands to the magnetic ordering should be weak~\cite{PhysRevB.78.195424}, producing a small electronic contribution to the response.
Rather, we expect a significant contribution would rely on the Khomskii polarization mechanism~\cite{khomskii2012}, wherein an electric dipole moment arises in tetrahedra that host spinon excitations pointing in the direction of the minority spin\footnote{ For example, the minority spin of a 3-in-1-out tetrahedron is the single outward-pointing spin. }.
In non-breathing pyrochlores, this cancels for adjacent spinon-antispinon pairs; with breathing, it need not. 
In the absence of an applied magnetic field, the staggered $Q$ order parameter corresponds to a spinon-antispinon zinc-blende crystal with no preferred orientation for the minority spins, and the associated Khomskii polarization cancels~\cite{khomskii2021electric}. 
An applied magnetic field would make certain spins preferentially `minority' for the existing zinc-blende crystal, which in turn leads to a linear electric polarization. 
A quantitative study of this microscopic mechanism is beyond the scope of the present paper.


\textit{Discussion} ---
We studied the dynamical interplay of axion and gauge fields in the context of quantum spin liquids (QSL), where they both appear as emergent phenomena. 
We identified a microscopic order parameter with the right symmetry to give rise to an emergent axion field.
This order is natural in quantum spin ice systems on breathing pyrochlore lattices~\cite{savary2016quantumBreath}, such as the QSL candidate ${\text{Ba}_{3}\text{Yb}_{2}\text{Zn}_{5}\text{O}_{11}}$~\cite{kimura2014experimental,PhysRevB.93.220407,PhysRevLett.116.257204,PhysRevB.98.054408}. 
The two main ingredients are the breaking of lattice inversion symmetry coexisting with low energy fluctuations of a scalar antiferromagnetic order parameter $Q$. 
From Eq.~\eqref{eqn:axionQSI}, a finite $\theta$ parameter requires $\langle Q \rangle \ne 0$.
Classically, such order can coexist with Coulomb liquid behavior, for example, in the fragmented phase of spin ice~\cite{brooks2014magnetic,lhotel2020fragmentation}, which has been recently observed in iridate materials~\cite{lefranccois2017fragmentation, PhysRevResearch.2.032073, pearce2021monopole}.
Our long-wavelength symmetry analysis suggests that the quantum limit of a fragmented spin liquid on breathing pyrochlore would be a $U(1)$ QSL phase with both a $\theta$-term and a dynamical axion.
Notice, however, that the dynamical part of the axion, $\phi$, comes from the amplitude fluctuations of $Q$, and is therefore present regardless of $\langle Q \rangle$. 

Unlike the axion electrodynamics studied in topological insulators and superconductors, and in Weyl semimetals~\cite{li2010dynamical, wu2016quantized, nenno2020axion,PhysRevB.87.134519,PhysRevB.87.161107,gooth2019axionic}, the quasiparticle content of $U(1)$ QSLs includes magnetic monopoles. 
In the presence of a finite $\theta$-term, the celebrated Witten effect induces the monopoles to acquire an electric charge, becoming dyons, 
%
%
%
which in turn 
leads to possible changes in the thermodynamic behavior of the system.
For instance, condensation of dyons can lead to distinct symmetry patterns in the neighboring ordered phases~\cite{cardy1982phase, cho2012dyon, PW220703544}.

Here, we focused instead on a dynamical axion, which modifies the inelastic response and may be observed directly in neutron scattering.
Using a long-wavelength description, we investigated the behavior of the dynamical structure factor and demonstrated how it is qualitatively modified by the axion-photon continuum (see Fig.~\ref{fig:heatmap}).
Our results provide an avenue to observe signatures of emergent axion electrodynamics in a class of frustrated magnetic systems.
These signatures crucially depend on the presence of the underlying $U(1)$ gauge structure and therefore also provide direct evidence of QSL behavior.

Dynamical axions in our universe were first hypothesized over 40 years ago as a remedy for the strong $CP$ problem in the standard model~\cite{PhysRevLett.38.1440}.
Today they are the subject of intense experimental searches as candidates for dark matter~\cite{duffy2009axions}, yet to be observed~\cite{graham2015experimental}.
In condensed matter systems, while pseudoscalar collective modes are not so elusive, their coupling to a fully internal electrodynamics is not readily available.
In this paper, we demonstrated how one can realize and access an emergent axion electrodynamics in the context of quantum spin liquids.
Speculatively, this provides a test bed for high-energy physics that cannot be currently probed directly due to limitations of experiments or, more dramatically, limitations in the content of the universe itself.


\textit{Acknowledgements} --- 
The authors are grateful to Anushya Chandran, Pieter Claeys, Michael DeMarco, Eduardo Fradkin, Joseph Maciejko, Siddhardh Morampudi, and David Tong for useful discussions.
S.D.P. acknowledges support from The Winston Churchill Foundation of the United States through the Churchill Scholarship, and support from the Henry W. Kendall Fellowship. 
This work was supported in part by the Engineering and Physical Sciences Research Council (EPSRC) under grants EP/M007065/1, EP/P034616/1 and EP/T028580/1 (C.C. and S.D.P.). 
C.R.L. acknowledges support from the NSF through grant PHY-1752727 and the generous hospitality of the Aspen Center for Physics, which is supported by NSF grant PHY-1607611. 


\bibliographystyle{apsrev4-1}
\bibliography{references}

\begin{thebibliography}{73}
\bibitem [{foo()}]{footnoteTRSbreaking}%
  \BibitemOpen
  \href@noop {} {}\bibinfo {note} {On the pyrochlore lattice, all-out-all-in
  order ($Q \neq 0$) either has all spins on $A$ tetrahedra pointing outwards
  while all spins on $B$ tetrahedra point inwards, or vice versa. Unlike the
  more familiar N\'{e}el order on a square lattice, translation by primitive
  lattice vectors does not reverse the all-in-all-out order and time reversal
  symmetry cannot be restored by such a proper transformation. There is a space
  group symmetry which flips $Q$: mirror reflection ($\mathcal{M}$). However,
  $\mathcal{M}$ involves a large, improper, transformation of space and it does
  not coarse-grain trivially. The combination of $\mathcal{M}$ with time
  reversal is in fact a symmetry of the microscopic system which is consistent
  with the theta term in the continuum.}\BibitemShut {Stop}%
\end{thebibliography}


\pagebreak
\onecolumngrid
\pagebreak


\section{Supplemental Material}
\setcounter{page}{1}
\setcounter{figure}{0} 
\setcounter{equation}{0} 


\section{Electromagnetic Units}
Here we briefly explain the choice of electromagnetic units in the main text. For reference, in SI units the Maxwell Lagrangian is given by
\begin{equation}
\frac{\epsilon}{2}|\bm{e_{\mathrm{SI}}}|^{2} - \frac{1}{2\mu}|\bm{b_{\mathrm{SI}}}|^{2}
\end{equation}
where $\epsilon$ and $\mu$ are the effective dielectric constant and magnetic permeability, respectively. The emergent speed of light $c$ follows the standard relationship $c = 1/\sqrt{\epsilon\mu}$. 

We work in Heaviside-Lorentz units~\cite{page1932electromagnetic}, a type of CGS units where $\epsilon_{0}=1$ and the magnetic field is divided by $c$ so $\bm{e}$ and $\bm{b}$ have the same units. In these units, the Maxwell Lagrangian is indeed 
\begin{equation}
    \mathcal{L}_{\gamma} = \frac{1}{2}|\bm{e}|^{2} - \frac{1}{2}|\bm{b}|^{2} \, . 
\end{equation}
The electric field emitted from an electric charge, with charge $e$, is 
\begin{equation}
    \bm{e} = \frac{e}{4\pi |\bm{x}|^{2}}\bm{\hat{r}},
\end{equation} 
and the magnetic field from a magnetic monopole, where $g$ is its magnetic charge, is 
\begin{equation}
    \bm{b} = \frac{g}{4\pi c|\bm{x}|^{2}}\bm{\hat{r}}.
\end{equation}
Futhermore, working with $\hbar=1$, in these units the fine structure constant is given by 
\begin{equation}
    \alpha = \frac{e^{2}}{4\pi c},    
\end{equation}
and Dirac's quantization is
\begin{equation}
    \frac{eg}{c^{2}} = 2\pi n,
\end{equation}
where $n\in\mathbb{Z}$.
Finally, we note that the emergent gauge field is $a^{\mu} = (a^{0},\bm{a})$, where $a^{0}$ is the electrostatic potential and $\bm{a}$ the vector potential. The electric and magnetic fields, in terms of $a^{0}$ and $\bm{a}$, are
\begin{align}
\label{eqn:eFieldDef0}\bm{e} &= -\grad a^{0} - \frac{1}{c}\partial_{t}\bm{a}, \\
\label{eqn:bFieldDef0}\bm{b} &= \grad\cross\bm{a},
\end{align}
where we do not include magnetic monopole contributions in $\bm{b}$ (see Eq.~\eqref{eqn:bFieldDef} further below).


\section{(Breathing) Pyrochlore Lattice and Local Easy Axes}
\begin{figure}[t]
\centering
    \includegraphics[width=\columnwidth]{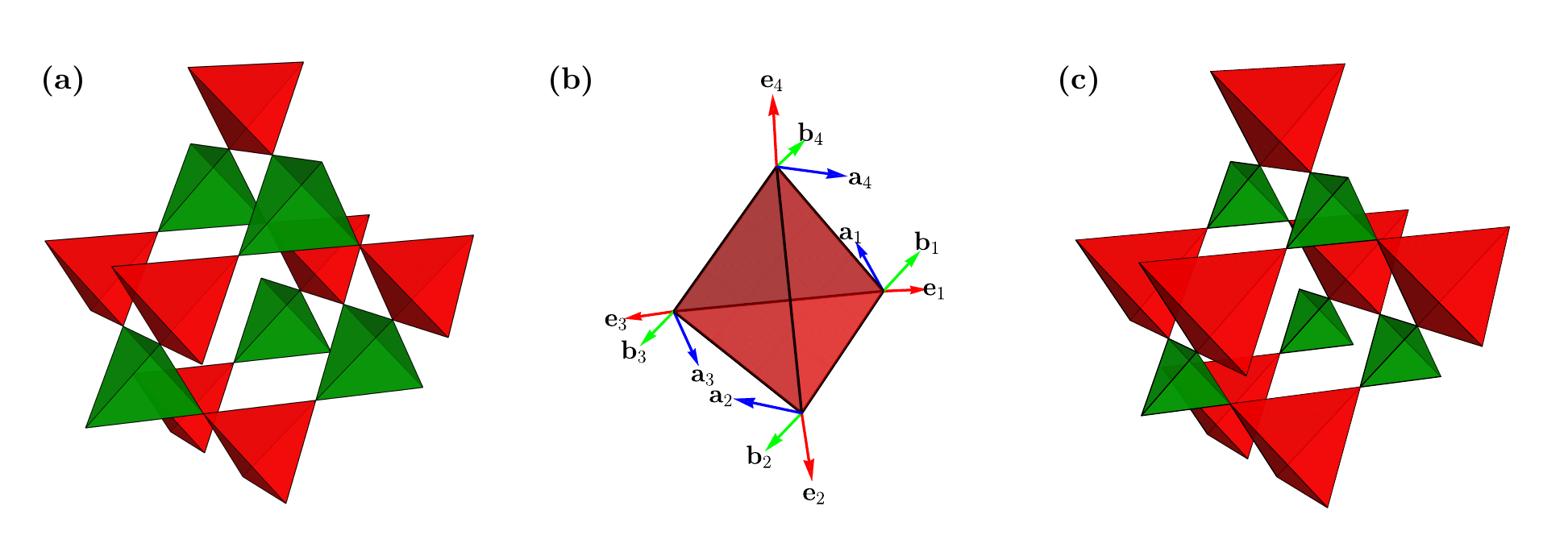}
    \caption{
    The pyrochlore lattice is a prototypical setting for frustrated magnetism in $3+1$D.
    Panel (a) shows a chunk of the pyrochlore lattice with $A$ tetrahedra colored in red and $B$ tetrahedra in green. 
    The four neighboring tetrahedra of an $A$ ($B$) tetrahedron are all $B$ ($A$) tetrahedra.
    %
    %
    Panel (b) shows an isolated $A$ tetrahedron with the four conventional local axes shown for each corner. 
    The $(\bm{\hat{a}}_{n},\bm{\hat{b}}_{n},\bm{\hat{e}}_{n})$ vectors are defined in Eq.~\eqref{eqn:localAxes}.
    %
    %
    (c) Breathing anisotropy refers to the $A$ and $B$ tetrahedra in the pyrochlore lattice becoming unequal in size. 
    The breathing pyrochlore lattice --- the pyrochlore lattice with breathing anisotropy --- is shown, where the $B$ tetrahedra have become smaller than $A$ tetrahedra.
    }
    \label{fig:pyroFig}
\end{figure}

The pyrochlore lattice is a network of corner-sharing tetrahedra that form a bipartite (diamond) lattice, as illustrated in Fig.~\ref{fig:pyroFig}(a) where sublattice $A$ and $B$ are shown in red and green, respectively.
In pyrochlore magnets, spin degrees of freedom reside on the corners of the tetrahedra, and in spin ice they are restricted to point directly into or out of each tetrahedron.
Because of the easy-axis (Ising) anisotropy, the spin operators are most conveniently defined in a local frame where a positive $z$-component of a spin points towards the center of a $B$ tetrahedron, along the local $[1,1,1]$ direction. 
Indeed, consider an $A$ tetrahedron centered at $\bm{x}$, whose four corners are labelled by $n=1,2,3,4$. 
In terms of the local frame, the spin vector $\bm{\mu}$ is given by
\begin{equation}\label{eqn:globalSpin}
\bm{\mu}(\bm{x},n) = 
S^{(x)}_{\bm{x},n}\bm{\hat{a}}_{n} + 
S^{(y)}_{\bm{x},n}\bm{\hat{b}}_{n} + 
S^{(z)}_{\bm{x},n}\bm{\hat{e}}_{n}
\, ,
\end{equation}
where the orthonormal vectors $(\bm{\hat{a}}_{i},\bm{\hat{b}}_{i},\bm{\hat{e}}_{i})$ defining the local easy axes are given by~\cite{savary2012coulombic}
\begin{equation}\label{eqn:localAxes}
\begin{array}{l}
\bm{\hat{a}}_{1} = \left( \dfrac{-2}{\sqrt{6}}, \dfrac{1}{\sqrt{6}}, \dfrac{1}{\sqrt{6}} \right),\vspace{5pt}  \\
\bm{\hat{a}}_{2} = \left( \dfrac{-2}{\sqrt{6}}, \dfrac{-1}{\sqrt{6}}, \dfrac{-1}{\sqrt{6}} \right), \vspace{5pt} \\
\bm{\hat{a}}_{3} = \left( \dfrac{2}{\sqrt{6}}, \dfrac{1}{\sqrt{6}}, \dfrac{-1}{\sqrt{6}} \right),  \vspace{5pt} \\
\bm{\hat{a}}_{4} = \left( \dfrac{2}{\sqrt{6}}, \dfrac{-1}{\sqrt{6}}, \dfrac{1}{\sqrt{6}} \right),\vspace{5pt}
\end{array}
\quad\quad
\begin{array}{l}
\bm{\hat{b}}_{1} = \left( 0, \dfrac{-1}{\sqrt{2}}, \dfrac{1}{\sqrt{2}} \right), \vspace{5pt} \\
\bm{\hat{b}}_{2} = \left( 0, \dfrac{1}{\sqrt{2}}, \dfrac{-1}{\sqrt{2}} \right),  \vspace{5pt}\\
\bm{\hat{b}}_{3} = \left( 0, \dfrac{-1}{\sqrt{2}}, \dfrac{-1}{\sqrt{2}} \right), \vspace{5pt} \\
\bm{\hat{b}}_{4} = \left( 0, \dfrac{1}{\sqrt{2}}, \dfrac{1}{\sqrt{2}} \right), \vspace{5pt} 
\end{array}
\quad\quad
\begin{array}{l}
\bm{\hat{e}}_{1} = \left( \dfrac{1}{\sqrt{3}}, \dfrac{1}{\sqrt{3}}, \dfrac{1}{\sqrt{3}} \right), \vspace{5pt}\\
\bm{\hat{e}}_{2} = \left( \dfrac{1}{\sqrt{3}}, \dfrac{-1}{\sqrt{3}}, \dfrac{-1}{\sqrt{3}} \right), \vspace{5pt} \\
\bm{\hat{e}}_{3} = \left( \dfrac{-1}{\sqrt{3}}, \dfrac{1}{\sqrt{3}}, \dfrac{-1}{\sqrt{3}} \right), \vspace{5pt} \\
\bm{\hat{e}}_{4} = \left( \dfrac{-1}{\sqrt{3}}, \dfrac{-1}{\sqrt{3}}, \dfrac{1}{\sqrt{3}} \right). \vspace{5pt}
\end{array}
\end{equation}
An $A$ tetrahedron with these four local coordinate systems is shown in Fig.~\ref{fig:pyroFig}(b).
In the main text, we focused on the magnetic moment part of the total spin vector. For simplicity, we used a single index $l \equiv (\bm{x},n)$ for the spin (pyrochlore) sites, and we further considered the coarse grained, continuum limit: $\bm{S}_{l} \to \bm{S}(\bm{x})$. 

As discussed in this work, in order to realise an emergent axion electrodynamics, we need a breathing pyrochlore lattice. 
Under such anisotropy, the $A$ and $B$ tetrahedra are of unequal size and the inversion symmetry about a site is broken, which reduces the pyrochlore lattice space group, $F d \overline{3} m$, down to $F \overline{4} 3 m$. 
Fig.~\ref{fig:pyroFig}(c) shows the breathing pyrochlore lattice with the $A$ tetrahedra (in red) larger than the $B$ tetrahedra (in green). 


\section{Symmetry Properties of the Emergent Fields in Quantum Spin Ice}\label{sec:sym}
Here we discuss the symmetry properties of the emergent gauge field, $a^{\mu}$, and axion field, $\phi$, which in turn derive from the combined symmetry properties of the microscopic spin-$1/2$ moments and of the lattice. 
We consider quantum spin ice (QSI), where spins reside on the pyrochlore lattice (see Fig.~\ref{fig:pyroFig}), and follow the notation $S^{(z)}_{l} = 1/2$ for a spin at site $i$ pointing into a $B$ tetrahedron (and $S^{(z)}_{l} = -1/2$ for a spin pointing into an $A$ tetrahedron).

We focus on spin-$1/2$ degrees of freedom realized by magnetic doublets with the following properties under time-reversal, $\mathcal{T}$, and inversion, $\mathcal{I}$, about a pyrochlore lattice site: 
\begin{align}
    \mathcal{T}:&\quad S^{(z)}_{l}(t)\to -S^{(z)}_{l}(-t),\\
    \mathcal{I}:&\quad \bm{S}_{l}(t) \to \hspace{8pt}\bm{S}_{\mathcal{I}(l)}(t).
\end{align}
These properties are satisfied by Kramers, non-Kramers, and dipolar–octupolar doublets, which make up all of the possible magnetic doublets in QSI candidate materials~\cite{rau2019frustrated}. 
They encompass the Kramers doublets carried by the Yb$^{3+}$ ions in Ba$_{3}$Yb$_{2}$Zn$_{5}$O$_{11}$, which is the quantum breathing pyrochlore magnet discussed in the main text.

The emergent gauge potential on the lattice is canonically conjugate to $S^{(z)}$, thus satisfying $[a_{l}(t),S^{(z)}_{l'}(t')] = i\delta_{ll'}\delta(t-t')$.
We can use this canonical commutation relation to find the symmetry properties of $a_{l}(t)$.
Requiring that $a_{l}(t)$ transforms in such a way that the commutation relations are left unchanged -- up to an constant shift which can be removed by a gauge transformation -- the emergent gauge potential transforms as
\begin{align}
    \mathcal{T}:&\quad a_{l}(t)\to a_{l}(-t),\label{eqn:Tatrans}\\
    \mathcal{I}:&\quad a_{l}(t) \to a_{\mathcal{I}(l)}(t).
\end{align}
For Eq.~\eqref{eqn:Tatrans}, we used that $\mathcal{T}$ is an anti-unitary operator that changes the sign of the imaginary unit $i$ in the commutation relation. 
From here, we can then connect the symmetry properties to the electric and magnetic field variables using the lattice definitions of the gradient and curl based on discrete differential forms. 
However, it is more convenient instead to first coarse-grain the vector potential and work with these differential operators in the continuum:
\begin{align}
    \mathcal{T}:&\quad \bm{a}(\bm{x},t)\to \bm{a}(\bm{x},-t)
    \, , \\
    \mathcal{I}:&\quad \bm{a}(\bm{x},t) \to \bm{a}(-\bm{x},t)
    \, . 
\end{align}

Taking into account that we have a compact gauge potential and thus magnetic monopoles, $\bm{e}$ and $\bm{b}$ are related to the emergent gauge field $a^{\mu} = (a^{0},\bm{a})$ by the relationships
\begin{align}
\label{eqn:eFieldDef}\bm{e} &= -\grad a^{0} - \frac{1}{c}\partial_{t}\bm{a}, \\
\label{eqn:bFieldDef}\bm{b} &= \grad\cross\bm{a} + \bm{n},
\end{align}
where $\grad\cdot\bm{n}$ equals the magnetic monopole magnetic charge $g$. 
From these definitions, we see that the electric field transforms like $\partial_{t}\bm{a}$ and the magnetic field transforms like $\grad\cross\bm{a}$. 
Therefore, under time-reversal the emergent electric and magnetic fields transform as
\begin{align}
\mathcal{T}:&\quad\bm{e}(\bm{x},t)\to -\bm{e}(\bm{x},-t) \, , \\
\mathcal{T}:&\quad\bm{b}(\bm{x},t)\to\hspace{8pt}\bm{b}(\bm{x} \, ,-t),
\end{align}
and under inversion they transform as
\begin{align}
\mathcal{I}:&\quad\bm{e}(\bm{x},t)\to \hspace{8pt}\bm{e}(-\bm{x},t) \, , \\
\mathcal{I}:&\quad\bm{b}(\bm{x},t)\to -\bm{b}(-\bm{x},t) \, .
\end{align}
We see that the emergent electric and magnetic fields transform oppositely under $\mathcal{T}$ and $\mathcal{I}$ when we compare them to the external electric and magnetic fields. 
However, $\bm{e}\cdot\bm{b}$ transforms in the same way in both emergent and external electromagnetism. 
In fact, because $\bm{e}\cdot\bm{b}$ transforms like $(\partial_{t}\bm{a}\cdot \grad\cross\bm{a})$, it is always a time-reversal odd pseudoscalar, regardless of how the gauge field transforms. 
Additionally, we see that $\grad a^{0}$ transforms like $\bm{e}$ and thus $a^{0}$ is time-reversal odd pseudoscalar. 

From the symmetry properties of the emergent electric and magnetic fields, we find the symmetry properties of the electric and magnetic charge density, $\rho_{e}(\bm{x},t)$ and $\rho_{m}(\bm{x},t)$ respectively. 
Under time-reversal, these charge densities transform as
\begin{align}
\mathcal{T}:&\quad\rho_{e}(\bm{x},t)\to -\rho_{e}(\bm{x},-t) \, , \\
\mathcal{T}:&\quad\rho_{m}(\bm{x},t)\to\hspace{4pt}\rho_{m}(\bm{x},-t) \, ,
\end{align}
and under inversion they transform as
\begin{align}
\mathcal{I}:&\quad\rho_{e}(\bm{x},t)\to -\rho_{e}(-\bm{x},t) \, , \\
\mathcal{I}:&\quad\rho_{m}(\bm{x},t)\to \hspace{4pt}\rho_{m}(-\bm{x},t) \, .
\end{align}
We see that $\rho_{m}(\bm{x},t)$ is a time-reversal even scalar field while $\rho_{e}(\bm{x},t)$ is a time-reversal odd pseudoscalar.
One may note that $\rho_{e}(\bm{x},t)$ has the required symmetry properties to interact with emergent photons as an emergent axion.
However, because it is the divergence of a vector field, it fails to satisfy the requirement of having well-defined local fluctuations that survive coarse-graining. 
As for other symmetries, the coarse-grained fields simply transform as vectors/pseudovectors or scalars/pseudoscalars.

We now discuss the symmetry properties of the order parameters $Q$ and $\Delta V$ defined in Eq.~\eqref{eqn:Qop} and~\eqref{eqn:deltaVop}, respectively, in the main text. 
The order parameter $Q$ can be written in terms of $S^{(z)}$ using the definition of the lattice divergence
\begin{equation}
    \operatorname{div}_{t}S^{(z)} = (-1)^{t} \sum_{l\in t}S^{(z)}_{l} \, ,
\end{equation}
where the sum is over the pyrochlore lattice sites, $l$, making up the four corners of tetrahedron $t$. 
Plugging this into Eq.~\eqref{eqn:Qop} yields
\begin{equation}
    Q_{t} = \sum_{l\in t}S^{(z)}_{l} \, .
\end{equation}
From the symmetry properties of $S^{(z)}$, it is straight-forward to see that $Q$ transforms under $\mathcal{T}$ and $\mathcal{I}$ as
\begin{align}
\mathcal{T}:&\quad Q_{t}(t)\to -Q_{t}(-t) \, , \\
\mathcal{I}:&\quad Q_{t}(t)\to \hspace{8pt}Q_{\mathcal{I}(t)}(t) \, .
\end{align}
Using the fact that the lattice remains unchanged under $\mathcal{T}$, $\Delta V_{l}$ clearly transforms as
\begin{equation}
\mathcal{T}:\quad \Delta V_{l}(t) \to \Delta V_{l}(-t) \, .
\end{equation}
As for inversion, because $\mathcal{I}$ takes every $A$ ($B$) tetrahedron to a $B$ ($A$) tetrahedron, we have that
\begin{equation}
\mathcal{I}:\quad \Delta V_{l}(t) \to -\Delta V_{\mathcal{I}(l)}(t) \, .
\end{equation}
Upon coarse-graining the order parameters, in the continuum limit the fields $Q(\bm{x},t)$ and $\Delta V(\bm{x},t)$ transform as
\begin{align}
\mathcal{T}:&\quad Q(\bm{x},t)\to -Q(\bm{x},-t)\quad\quad \Delta V(\bm{x},t)\to\hspace{8pt}\Delta V(\bm{x},-t) \, , \\
\mathcal{I}:&\quad Q(\bm{x},t)\to \hspace{8pt}Q(-\bm{x},t)\quad\quad \Delta V(\bm{x},t)\to -\Delta V(-\bm{x},t) \, .
\end{align}
Thus, as stated in the main text, the product is a time-reversal odd pseudoscalar:
\begin{align}
\mathcal{T}:&\quad \left[Q(\bm{x},t)\right]\left[\Delta V(\bm{x},t)\right]\to -\left[Q(\bm{x},-t)\right]\left[\Delta V(\bm{x},-t)\right] \, ,\\
\mathcal{I}:&\quad \left[Q(\bm{x},t)\right]\left[\Delta V(\bm{x},t)\right]\to -\left[Q(-\bm{x},t)\right]\left[\Delta V(-\bm{x},t)\right] \, .
\end{align}


\section{Construction of the effective action}
In this section, we present the construction of $S_{\text{eff}}$ given in the main text. There are three long-wavelength dynamical fields which do not coarse grain to zero in the problem: the axion field $\phi$, the internal electric field $\bm{e}$, and the internal magnetic field $\bm{b}$. From SM Sec.~\ref{sec:sym}, under time reversal and inversion, these fields transform as
\begin{align*}
    &\underline{\text{Time Reversal}}\hspace{100pt}\underline{\text{Inversion}}\\
    \phi(t&,\bm{x})\to-\phi(-t,\bm{x})\hspace{60pt}\phi(t,\bm{x})\to-\phi(t,-\bm{x})\\
    \bm{e}(t&,\bm{x})\to-\bm{e}(-t,\bm{x})\hspace{60pt}\bm{e}(t,\bm{x})\to \bm{e}(t,-\bm{x})\\
    \bm{b}(t&,\bm{x})\to\bm{b}(-t,\bm{x})\hspace{70pt}\bm{b}(t,\bm{x})\to-\bm{b}(t,-\bm{x}) \, .
\end{align*}
From these fields, we construct local, analytic, rotationally invariant terms
\begin{align*}
    \underline{\text{zero}:}&\quad \phi,\\
    \underline{\text{one}:}&\quad \bm{e},~\bm{b},~\partial_t\phi,~\bm{\nabla}\phi\\
    \underline{\text{two}:}&\quad \bm{\nabla}\cdot\bm{e},~\bm{\nabla}\cdot\bm{b},~\bm{\nabla}\cross\bm{e},~\bm{\nabla}\cross\bm{b},~\partial_t\bm{e},~\partial_t\bm{b},~\partial_t^2\phi,~\bm{\nabla}^2\phi,~\partial_t\bm{\nabla}\phi
\end{align*}
which are organized by their number of spacetime derivatives. Using these, we can then find the analytic, time-reversal invariant scalars that may appear in the effective action. The terms containing up to two spacetime derivatives and up to three fields, which involve the emergent photon are:
\begin{align*}
&\bm{e} \cdot \bm{e} \quad\quad\quad\quad \phi (\bm{e} \cdot \bm{b}) \quad\quad\quad\quad \bm{e} \cdot \bm{\nabla} \phi \quad\quad\quad\quad \phi^2 (\bm{\nabla} \cdot \bm{b}) \\
&\bm{b} \cdot \bm{b} \quad\quad\quad\quad \phi (\bm{\nabla} \cdot \bm{e}) \quad\quad\quad\quad \bm{\nabla} \cdot \bm{b} \quad\quad\quad\quad \phi (\bm{b} \cdot \bm{\nabla} \phi) \, .
\end{align*}

This collection of terms can be further reduced. Firstly, assuming the magnetic charge is conserved, ${\int\bm{\nabla}\cdot\bm{b} = 0}$. Thus, the term ${\bm{\nabla}\cdot\bm{b}}$ will not appear by itself in the Lagrangian. Secondly, up to a total derivative ${\boldsymbol{e} \cdot \boldsymbol{\nabla} \phi = \phi(\boldsymbol{\nabla} \cdot e)}$, and without loss of generality we can drop the term ${\boldsymbol{e} \cdot \boldsymbol{\nabla} \phi}$. Furthermore, the Gauss-Witten law, Eq.~\eqref{eqn:axionGaussLaw} in the main text 
\begin{equation*}
    \bm{\nabla} \cdot \bm{e}=-\frac{\alpha}{\pi} \phi (\bm{\nabla} \cdot \bm{b})-\frac{\alpha}{\pi} ~\bm{b} \cdot \bm{\nabla} \phi \, ,
\end{equation*}
is enforced at the path integral level as well. Thus, the term ${\phi(\grad\cdot \bm{e})}$ is equivalent to ${\phi^2 (\bm{\nabla} \cdot \bm{b})}$ and ${\phi (\bm{b} \cdot \bm{\nabla} \phi)}$. Taking all of these considerations into account, the list of allowed terms reduces to
\begin{equation*}
    \begin{aligned}
        &\bm{e} \cdot \bm{e} \quad\quad\quad\quad \phi (\bm{e} \cdot \bm{b}) \quad\quad\quad\quad  \phi^2 (\bm{\nabla} \cdot \bm{b}) \\
&\bm{b} \cdot \bm{b} \quad\quad\quad\quad  \phi (\bm{b} \cdot \bm{\nabla} \phi ) \, .
    \end{aligned}
\end{equation*}
Three of these terms precisely make up the Maxwell Lagrangian and Axion term. For simplicity, the two other terms are not included in our calculations in the main text; indeed, as we shall explain hereafter, they do not alter our results. 

Note how when $\phi$ is static, the two additional terms above vanish from the action (one is a boundary term and the other is zero). Thus, these terms only affect \emph{dynamical} processes involving the axion. They both contribute to an additional spectral continuum where an emergent photon excites two axions. This two-particle continuum is different from the one studied in the main text, which is the photon-axion continuum. Furthermore, the threshold energy of this two particle continuum scales with twice the axion gap, which is larger than the threshold for the photon-axion continuum and, therefore, will not affect the latter up to much higher energies.

At higher orders in perturbation theory, these terms (or other irrelevant perturbations) could affect quantitative details of the axion-photon spectral continuum but they do not affect the primary result presented in the main text, illustrated by Eq.~\eqref{eqn:fThBehavoir}, namely the turn-on of the continuum. Near the threshold, the nonanalyticity governing the turn-on of the spectral function (i.e., the exponents in Eq.~\eqref{eqn:fThBehavoir}) are therefore expected to be robust. This is a general feature of threshold spectra which are typically governed by the onset of the density of asymptotic (free particle) states along with general features of the local coupling (see, e.g., Ref.~\onlinecite{PhysRev.73.1002}).


\section{Dynamical Structure Factor Calculation Details}\label{sec:calc}

Here, we sketch the calculation of the dynamical structure factor (DSF) $\mathcal{F}^{\aind\bind}(\bm{k},\omega)$, with $\aind,\bind=x,y,z$, given by Eq.~\eqref{eqn:NSFdymSF} in the main text. 
We work at zero temperature with a real time path integral approach and use the notation $\langle\hspace{1pt}\cdots\rangle$ to refer to the real time-ordered vacuum expectation value.
According to linear response theory, the DSF $\mathcal{F}^{\aind\bind}(\bm{k},\omega)$ is given by the imaginary part of the dynamical spin susceptibility.
Relating this to the time-ordered correlators, we find
\begin{equation}\label{eqn:DSFintermsofCorrelator}
    \mathcal{F}^{\aind\bind}(\bm{k},\omega) = \operatorname{Im}\left[i \langle S^{\aind}(\bm{k},\omega)S^{\bind}(-\bm{k},-\omega)\rangle\right].
\end{equation}
Although we use relativistic notation where convenient (with metric $\eta = \operatorname{diag}(1,-1,-1,-1)$ and $c=1$), the theory is isotropic but not Lorentz invariant. 
Hence, the tensor symmetry analysis below refers to the rotational 3-tensor structure -- not the Minkowski 4-tensor structure, since the theory need not be covariant with respect to boosts.

We recall from main text Eq.~\eqref{eqn:SzAxion} that the long-wavelength spin vector field $\bm{S} = \bm{e} + \frac{\alpha}{\pi}\varphi \bm{b}$. 
Therefore,  
\begin{eqnarray}
&&\langle S^{\aind}(x)S^{\bind}(y)\rangle = \langle e^{\aind}(x)e^{\bind}(y)\rangle  
\\
&& \quad 
+\frac{\alpha}{\pi}\bigg[\theta\left(\langle e^{\aind}(x)b^{\bind}(y) \rangle + \langle b^{\aind}(x)e^{\bind}(y) \rangle\right) + \langle e^{\aind}(x)\phi(y)b^{\bind}(y)\rangle + \langle \phi(x) b^{\aind}(x)e^{\bind}(y)\rangle \bigg]
\nonumber \\
&& \quad 
+\frac{\alpha^{2}}{\pi^{2}}\bigg[\theta^{2}\langle b^{\aind}(x)b^{\bind}(y)\rangle + \langle b^{\aind}(x)\phi(x)b^{\bind}(y)\phi(y)\rangle 
+ \theta\left(\langle b^{\aind}(x)\phi(y)b^{\bind}(y)\rangle + \langle \phi(x)b^{\aind}(x)b^{\bind}(y)\rangle \right)\bigg]
\, ,
\nonumber 
\end{eqnarray}
where we use the notation $x \equiv (t,\bm{x})$.
As $\langle S^{\aind}(\bm{k},\omega)S^{\bind}(-\bm{k},-\omega)\rangle $ is a symmetric 2-tensor, the terms linearly proportional to $\theta$ must be symmetric 2-pseudotensors. 
The only symmetric 2-pseudotensors available to construct this correlator are $\theta \delta^{\aind\bind}$ and $\theta k^{\aind} k^{\bind}$. 
However, as the static $\theta$ term is a boundary term, it does not appear in any bulk correlation functions and therefore these terms must vanish.
With this simplification,
\begin{equation}\label{eqn:spinspinCorPosSpace}
\begin{aligned}
\langle S^{\aind}(x)S^{\bind}(y)\rangle =& \langle e^{\aind}(x)e^{\bind}(y)\rangle + \frac{\alpha}{\pi}\left[\langle e^{\aind}(x)\phi(y)b^{\bind}(y)\rangle + \langle \phi(x) b^{\aind}(x)e^{\bind}(y)\rangle \right]\\
&+\frac{\alpha^{2}}{\pi^{2}}\left[\theta^{2}\langle b^{\aind}(x)b^{\bind}(y)\rangle + \langle b^{\aind}(x)\phi(x)b^{\bind}(y)\phi(y)\rangle\right]
\, .
\end{aligned}
\end{equation}

The correlators follow from the low-energy effective action using the path integral
\begin{equation}
\mathcal{Z} = \int \mathcal{D}a \mathcal{D}\phi \exp\left[iS_{\mathrm{eff}}\right].
\end{equation}
Rewriting the effective action in the main text, Eq.~\eqref{eqn:action}, in terms of the gauge field $a$, we have 
\begin{equation}
S_{\mathrm{eff}} = \frac{1}{2}\int d^{4}x\left\{a^{\mu}\left(\eta_{\mu\nu}\partial_{\rho}\partial^{\rho} \right)a^{\nu}- J\phi\left(\partial_{t}^{2}-v^{2}\grad^{2} + \Delta^{2} \right)\phi + \frac{\alpha}{\pi}(\theta + \phi)\epsilon^{\mu\nu\rho\sigma}\partial_{\mu}a_{\nu}\partial_{\rho}a_{\sigma}\right\}.
\end{equation}
The first term is the Maxwell Lagrangian, where we used the Faddeev-Popov gauge fixing procedure, choosing to work in the Feynman gauge. 
The third term is the axion term, where $\epsilon^{\mu\nu\rho\sigma}$ is the totally anti-symmetric Levi-Civita symbol satisfying $\epsilon_{0123} = -\epsilon^{0123} = 1$.
From $S_{\mathrm{eff}}$, the free photon propagator in momentum space is given by
\begin{equation}
D_{0}^{\mu\nu}(k) = \frac{-i\eta^{\mu\nu}}{\omega^{2} - |\bm{k}|^{2}+i0^{+}}
\end{equation}
while the free axion propagator is
\begin{equation}
\Delta_{F}(k) = \frac{iJ^{-1}}{\omega^{2} - v^{2}|\bm{k}|^{2} - \Delta^{2} + i0^{+}},
\end{equation}
where we use the notation $k\equiv(\omega,\bm{k})$.

We proceed by perturbatively evaluating the correlation functions in Eq.~\eqref{eqn:spinspinCorPosSpace} to find $\langle S^{\aind}(x)S^{\bind}(y)\rangle$ to leading loop order. 
Schematically, we evaluate in the following approximation 
\begin{align}
\label{eqn:diagramcorrelator}
\langle S^{\aind}S^{\bind}\rangle&=\raisebox{-1.1ex}{\includegraphics[width=0.9\columnwidth]{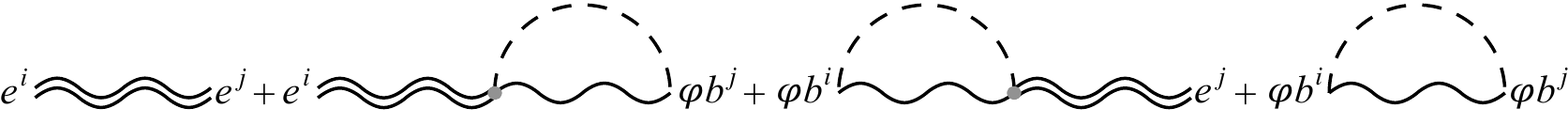}}
\end{align}
where the renormalized photon propagator $D^{\mu\nu}$ is given by solving the Schwinger-Dyson equation,
\begin{align}\label{eqn:SDeqn}
    \includegraphics[]{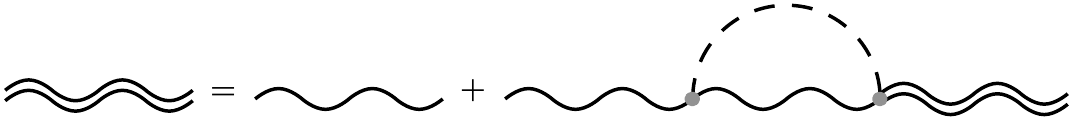}
\end{align}
The vacuum polarization tensor (the photon self-energy), $\Pi_{\rho\sigma}(k)$ is given by
\begin{equation}\label{eqn:polarTen}
i\Pi^{\mu\nu}(k) = -\frac{\alpha^{2}}{\pi^{2}}\int\frac{d^{4}p}{(2\pi)^{4}}\epsilon\indices{^{\mu}_{\alpha\tau\delta}}\epsilon\indices{^{\nu}_{\beta\gamma\lambda}}\hspace{3pt}k^{\alpha}k^{\beta}p^{\tau}p^{\gamma}\Delta_{F}(k-p)D_{0}^{\lambda\delta}(p).
\end{equation}
Because $\Pi_{\rho\sigma}(k)$ is transverse --- $k_{\mu}\Pi^{\mu\nu} = 0$ --- and our theory is rotationally invariant, we can write the polarization tensor as
\begin{equation}
\Pi^{\mu\nu} = \Pi_{1}(\omega,|\bm{k}|)P_{1}^{\mu\nu} + \Pi_{2}(\omega,|\bm{k}|)P_{2}^{\mu\nu},
\end{equation}
where we define the orthogonal projectors --- $ (P_{\ell})\indices{^{\mu}_{\sigma}} (P_{\ell^\prime})^{\sigma\nu} = \delta_{\ell\ell^\prime} (P_{\ell})^{\mu\nu}$, where $\ell,\ell^\prime=1,2$ ---
\begin{align}
    P^{\mu\nu}_{1} &= 
    \begin{pmatrix}
    0 && \bm{0}\\
    \bm{0} && \frac{k^{\aind}k^{\bind}}{|\bm{k}|^{2}} - \delta^{\aind\bind}
    \end{pmatrix}\\
    P^{\mu\nu}_{2} &= \frac{1}{|\bm{k}|^{2} - \omega^{2}}
    \begin{pmatrix}
    |\bm{k}|^{2} && \omega\bm{k}\\
    \omega\bm{k} && \omega^{2}\frac{k^{\aind}k^{\bind}}{|\bm{k}|^{2}}
    \end{pmatrix},
\end{align}
and $\Pi_{1}$ and $\Pi_{2}$ are rotational scalars. Expanding the dressed photon-propagator $D^{\mu\nu}$ in terms of these projectors and $P_{3}^{\mu\nu} = \eta^{\mu\nu} - P_{1}^{\mu\nu} - P_{2}^{\mu\nu}$, Eq.~\eqref{eqn:SDeqn} yields
\begin{equation}\label{eqn:renormalPho}
    D^{\mu\nu} = \frac{-iP_{1}^{\mu\nu}}{\omega^{2} - |\bm{k}|^{2} - \Pi_{1} + i0^{+}} + \frac{-iP_{2}^{\mu\nu}}{\omega^{2} - |\bm{k}|^{2} - \Pi_{2} + i0^{+}} + \frac{-iP_{3}^{\mu\nu}}{\omega^{2} - |\bm{k}|^{2} + i0^{+}}.
\end{equation}
The weights $\Pi_{1}$ and $\Pi_{2}$ can be found by considering the trace of $\Pi$, yielding
\begin{equation}
    \Pi\indices{^{\mu}_{\mu}} = 2\Pi_{1} + \Pi_{2},
\end{equation}
and the $00$ component
\begin{equation}
    \Pi^{00} = \frac{|\bm{k}|^{2}}{|\bm{k}|^{2} - \omega^{2}}\Pi_{2}.
\end{equation}
Solving for $\Pi^{00}$ and $\Pi\indices{^{\mu}_{\mu}}$ using Eq.~\eqref{eqn:polarTen}, we find the integral forms
\begin{align}
    i\Pi_{1} &= \frac{\alpha^{2}}{J\pi^{2}}\int \frac{d^{4}p}{(2\pi)^{4}}\bigg((\omega^{2}-|\bm{k}|^{2})(p_{0}^{2} - |\bm{p}|^{2}) - (p_{0}\omega - \bm{k}\cdot\bm{p})\nonumber\\
    &\hspace{100pt}+\frac{1}{2}(\omega^{2}-|\bm{k}|^{2})|\bm{p}|^{2}(1-
\cos^{2}(\theta))\bigg)\Delta_{F}(k-p)D_{0}(p),\\
i\Pi_{2} &= \frac{\alpha^{2}}{J\pi^{2}}\int \frac{d^{4}p}{(2\pi)^{4}}(\omega^{2}-|\bm{k}|^{2})|\bm{p}|^{2}(\cos^{2}(\theta) - 1)\Delta_{F}(k-p)D_{0}(p),
\end{align}
where $\theta$ is the polar angle between $\bm{k}$ and $\bm{p}$ and $D_{0}(p)$ is defined by $D^{\mu\nu}_{0}(p) \equiv \eta^{\mu\nu}D_{0}(p)$. 

Given the photon propagator and the axion propagator, we can return to Eq.~\eqref{eqn:spinspinCorPosSpace} and calculate each term in momentum space using the definitions $e^{\aind} = \partial^{\aind}a^{0} - \partial^{0}a^{\aind}$ and $b^{\aind} = \epsilon^{\aind\bind\cind}\partial_{\bind} a^{\cind}$. 
Under the approximation where only external photon propagators are dressed, Eq.~\eqref{eqn:diagramcorrelator}, we find  
\begin{align}
\langle e^{\aind}(k)e^{\bind}(-k)\rangle &= k^{\aind}k^{\bind}D^{00} - \omega k^{\bind} D^{\aind 0} - \omega k^{\aind} D^{\bind 0} + \omega^{2}D^{\aind\bind},\\
\langle b^{\aind}(k)b^{\bind}(-k)\rangle &= \epsilon^{\aind mn}\epsilon^{\bind\ell p}k^{m}k^{\ell}D^{np},\\
\langle [b^{\aind}\phi](k)[b^{\bind}\phi](-k)\rangle &= \int \frac{d^{4}p}{(2\pi)^{4}}(p^{\aind}p^{\bind} - \delta |\bm{p}|^{2})D_{0}(p)\Delta_{F}(k-p),\\
\langle e^{\aind}(k)[\phi b^{\bind}](-k)\rangle + \langle [\phi b^{\aind}](k)e^{\bind}(-k)\rangle &= i\frac{\alpha}{\pi}\epsilon^{\tau\lambda\rho n}\left[ \epsilon^{0\bind\ell n}\left(k^{\aind} D^{0}_{\lambda} - \omega D^{\aind}_{\lambda}\right) + \aind\leftrightarrow \bind\right]k_{\tau}\nonumber\\
&\hspace{40pt}\times\int\frac{d^{4}p}{(2\pi)^{4}}p_{\rho}p^{\ell}\Delta_{F}(k-p)D_{0}(p).
\end{align}
Plugging in the expression for the dressed photon propagator, Eq.~\eqref{eqn:renormalPho}, and simplifying yields
\begin{equation}\label{eqn:eeCor1}
    \langle e^{\aind}(k)e^{\bind}(-k)\rangle = \left(\frac{i\omega^{2}}{\omega^{2} - |\bm{k}|^{2} -\Pi_{1} + i0^{+}}\right)P^{\aind\bind}_{T} +\left(\frac{i(\omega^{2}-|\bm{k}|^{2})}{\omega^{2} - |\bm{k}|^{2} -\Pi_{2} + i0^{+}}\right)P^{\aind\bind}_{L},
\end{equation}
\begin{equation}\label{eqn:bbCor1}
    \langle b^{\aind}(k)b^{\bind}(-k)\rangle = \left(\frac{i|\bm{k}|^{2}}{\omega^{2} - |\bm{k}|^{2} -\Pi_{1} +i0^{+}}\right)P_{T}^{\aind\bind},
\end{equation}
\begin{equation}\label{eqn:bphibphiCor1}
    \langle [b^{\aind}\phi](k)[b^{\bind}\phi](-k)\rangle = \int \frac{d^{4}p}{(2\pi)^{4}}(p^{\aind}p^{\bind} - \delta |\bm{p}|^{2})D_{0}(p)\Delta_{F}(k-p),
\end{equation}
\begin{equation}\label{eqn:ebphiCor1}
    \begin{aligned}
        &\langle e^{\aind}(k)[\phi b^{\bind}](-k)\rangle + \langle [\phi b^{\aind}](k)e^{\bind}(-k)\rangle = \frac{\alpha}{\pi}\int\frac{d^{4}p}{(2\pi)^{4}}\bigg[\frac{\left(1-\frac{\omega^{2}}{|\bm{k}|^{2}}\right)\left((k^{\aind}p^{\bind}+p^{\aind}k^{\bind})\bm{p}\cdot\bm{k}- 2k^{\aind}k^{\bind}|\bm{p}|^{2}\right)}{\omega^{2} - |\bm{k}|^{2} -\Pi_{2} + i0^{+}}\\
        &+\frac{2\omega\left(\delta^{\aind\bind}(|\bm{p}|^{2}\omega - \bm{k}\cdot\bm{p} p_{0}) - \frac{|\bm{p}|^{2}\omega}{|\bm{k}|^{2}}k^{\aind}k^{\bind} - \omega p^{\aind}p^{\bind}+(p_{0} - \frac{\bm{k}\cdot\bm{p}\omega}{|\bm{k}|^{2}})(\frac{k^{\aind}p^{\bind} + p^{\aind}k^{\bind}}{2})\right)}{\omega^{2} - |\bm{k}|^{2} -\Pi_{1} + i0^{+}}\bigg]D_{0}(p)\Delta_{F}(k-p) \, .
    \end{aligned}
\end{equation}
In Eqs.~\eqref{eqn:eeCor1} and~\eqref{eqn:bbCor1}, we have introduced the longitudinal and transverse projectors
\begin{align}
    P^{\aind\bind}_{L} &= \frac{k^{\aind}k^{\bind}}{|\bm{k}|^{2}}\\
    P^{\aind\bind}_{T} &= \delta^{\aind\bind} - \frac{k^{\aind}k^{\bind}}{|\bm{k}|^{2}}
\, , 
\end{align}
respectively.

Because all these expressions are symmetric 2-tensors, on symmetry grounds they must have the general form
\begin{equation}\label{eqn:tensorForm}
    M^{\aind\bind} = M^{L}P^{\aind\bind}_{L} + M^{T}P^{\aind\bind}_{T},
\end{equation}
where $M^{L}$ and $M^{T}$ are rotational scalars. 
Eqs.~\eqref{eqn:eeCor1} and~\eqref{eqn:bbCor1} are already in this form. 
For Eqs.~\eqref{eqn:bphibphiCor1} and~\eqref{eqn:ebphiCor1}, by considering the trace and the $33$ component of each expression, we can find integral expressions for the respective $M^{L}$ and $M^{T}$ weights. 
In what follows, we assume that the real part of the polarization tensor $\Pi$, which renormalizes the dispersion, is zero --- i.e., that the dispersion parameters are already renormalized. 
We consistently drop all other UV-divergent contributions.

To compute the DSF, we take the real part of Eqs.~(\ref{eqn:eeCor1}-\ref{eqn:ebphiCor1}) in the expression for Eq.~\eqref{eqn:tensorForm}. 
Computing the relevant loop integrals using the Cutkosky rules, we find that the structure factor has two different functional forms depending on the energy regime. 
For energies below the threshold energy
\begin{equation}
\omega_{\mathrm{th}} = 
\begin{cases}
\omega_{\phi}\left(\bm{k}\right) \quad &|\bm{k}| < k_{c}\\
\omega_{\gamma}\left(|\bm{k}|-k_{c}\right) + \omega_{\phi}\left(k_{c}\right) \quad &|\bm{k}| \geq k_{c}
\end{cases} 
\, ,
\end{equation}
where ${k_{c} = \Delta /(v\sqrt{v^{2}-1)}}$, the DSF is given by
\begin{equation}
    \mathcal{F}^{\aind\bind}\bigg|_{\omega<\omega_{\mathrm{th}}} = P^{\aind\bind}_{T}\left(\pi + \frac{\alpha^{2}\theta^{2}}{\pi}\right)\omega^{2}\delta(\omega^{2}-|\bm{k}|^{2}).
\end{equation}
Above the threshold energy, the DSF instead takes the form
\begin{equation}\label{eqn:DSFaboveThres}
    \begin{aligned}
    \left(\frac{\alpha^{2}}{J\pi^{2}}\right)^{-1}\mathcal{F}^{\aind\bind}\bigg|_{\omega>\omega_{\mathrm{th}}} &= P_{L}^{\aind\bind}\left(1 - \frac{(\omega^{2}-|\bm{k}|^{2})^{2}}{(\omega^{2}-|\bm{k}|^{2})^{2} + \alpha^{4}J^{-2}I_{2}^{2}}\right)M^{L}_{b\phi b \phi}\\ &\hspace{30pt}+ P_{T}^{\aind\bind}\left[\frac{I_{1}\pi^{2}\left(\omega^{2}+\frac{\alpha^{2}\theta^{2}}{\pi^{2}}|\bm{k}|^{2}\right)+(\omega^{2}-|\bm{k}|^{2}) M^{T}_{eb\phi}}{(\omega^{2}-|\bm{k}|^{2})^{2} + \alpha^{4}J^{-2}I_{1}^{2}}+M^{T}_{b\phi b\phi}\right].
    \end{aligned}
\end{equation}
In this expression, the quantities $I_{1}$, $I_{2}$,$M^{L}_{b\phi b \phi}$, $M^{T}_{b\phi b \phi}$, and $M^{T}_{e b \phi}$ are given by the integrals
\begin{align}
    I_{1} &= -\frac{1}{16 \pi^{3} v^{2} |\bm{k}|}\int_{p_{-}}^{p_{+}}dp \hspace{2pt}p^{2}\left(\frac{\omega^{2}-|\bm{k}|^{2}}{2}(1-f(p)^{2}) - (\omega - k f(p))^{2}\right),\\ 
    I_{2} &= -\frac{1}{16 \pi^{3} v^{2} |\bm{k}|}\int_{p_{-}}^{p_{+}}dp \hspace{2pt}p^{2}\left(\omega^{2} - |\bm{k}|^{2}\right)(f(p)^{2}-1),\\
    M^{L}_{b\phi b \phi} &= -\frac{1}{16 \pi v^{2} |\bm{k}|}\int_{p_{-}}^{p_{+}}dp \hspace{2pt}p^{2}\left(f(p)^{2}-1\right),\\ 
    M^{T}_{b\phi b \phi} &= -\frac{1}{16 \pi v^{2} |\bm{k}|}\int_{p_{-}}^{p_{+}}dp \hspace{2pt}p^{2}\left(\frac{-1-f(p)^{2}}{2}\right),\\ 
    M^{T}_{e b \phi} &= -\frac{1}{16 \pi v^{2} |\bm{k}|}\int_{p_{-}}^{p_{+}}dp \hspace{2pt}p^{2}\left(\omega^{2} - 2|\bm{k}| \omega f(p) + \omega^{2} f(p)^{2}\right) \, ,
\end{align}
where the function $f(p)$ is given by
\begin{equation}
    f(p) = \frac{v^{2}|\bm{k}|^{2} + \Delta^{2} + (p+p v - \omega)(-p + pv + \omega)}{2|\bm{k}|pv^{2}}
\end{equation}
and the bounds on the integrals are
\begin{align}
    p_{-} &= 
    \begin{cases}
       \dfrac{v^{2}|\bm{k}| - \omega - \sqrt{\Delta^{2} + v^{2}(|\bm{k}| - \Delta - \omega)(|\bm{k}| + \Delta - \omega)}}{v^{2}-1} & v > 1 \text{ and } \omega < \omega_{\varphi}(\bm{k})\\\\
       \dfrac{-v^{2}|\bm{k}| - \omega + \sqrt{\Delta^{2} + v^{2}(|\bm{k}| - \Delta + \omega)(|\bm{k}| + \Delta + \omega)}}{v^{2}-1} & \omega > \omega_{\varphi}(\bm{k})
    \end{cases},\\\nonumber\\
    p_{+} &= \dfrac{v^{2}|\bm{k}| - \omega + \sqrt{\Delta^{2} + v^{2}(|\bm{k}| - \Delta - \omega)(|\bm{k}| + \Delta - \omega)}}{v^{2}-1} \, .
\end{align}

Because the spin field is divergence free, the correlation function and thus the DSF should be purely transverse. 
We see that to leading order in $\alpha^{2}/J$, the longitudinal component of Eq.~\eqref{eqn:DSFaboveThres} indeed vanishes, leaving only the transverse part. However, at high-orders of $\alpha^{2}/J$ this is not the case, which is an unphysical byproduct of our finite order perturbative treatment. 
Therefore, we finally drop the longitudinal part to arrive at the expression for the DSF studied in the main text:
\begin{equation}
    \mathcal{F}^{\aind\bind}\bigg|_{\omega>\omega_{\mathrm{th}}} = \left(\frac{\alpha^{2}}{J\pi^{2}}\right)P_{T}^{\aind\bind}\left[\frac{I_{1}\pi^{2}\left(\omega^{2}+\frac{\alpha^{2}\theta^{2}}{\pi^{2}}|\bm{k}|^{2}\right)+(\omega^{2}-|\bm{k}|^{2}) M^{T}_{eb\phi}}{(\omega^{2}-|\bm{k}|^{2})^{2} + \alpha^{4}J^{-2}I_{1}^{2}}+M^{T}_{b\phi b\phi}\right] \, .
\end{equation}


\end{document}